# Robot-Assisted Mindfulness Practice: Analysis of Neurophysiological Responses and Affective State Change

Maryam Alimardani, Linda Kemmeren, Kazuki Okumura, and Kazuo Hiraki

*Abstract*— Mindfulness is the state of paying attention to the present moment on purpose and meditation is the technique to obtain this state. This study aims to develop a robot assistant that facilitates mindfulness training by means of a Brain-Computer Interface (BCI) system. To achieve this goal, we collected EEG signals from two groups of subjects engaging in a meditative vs. non-meditative human-robot interaction (HRI) and evaluated cerebral hemispheric asymmetry, which is recognized as a well-defined indicator of emotional states. Moreover, using self-reported affective states, we strived to explain asymmetry changes based on pre- and post-experiment mood alterations. We found that unlike earlier meditation studies, the fronto-central activations in alpha and theta frequency bands were not influenced by robot-guided mindfulness practice, however there was a significantly greater right-sided activity in the occipital gamma band of Meditation group, which is attributed to increased sensory awareness and open monitoring. In addition, there was a significant main effect of Time on participant's self-reported affect, indicating an improved mood after interaction with the robot regardless of the interaction type. Our results suggest that EEG responses during robot-guided meditation hold promise in real-time detection and neurofeedback of mindful state to the user, however the experienced neurophysiological changes may differ based on the meditation practice and recruited tools. This study is the first to report EEG changes during mindfulness practice with a robot. We believe that our findings driven from an ecologically valid setting, can be used in development of future BCI systems that are integrated with social robots for health applications.

## I. INTRODUCTION

Mindfulness is the state of paying attention to the present moment on purpose [1] and meditation is the technique to achieve this state. Mindfulness meditation is widely practiced for reduction of stress and anxiety, regulation of emotions and treatment of depression. Many studies have already indicated potential benefits of mindfulness practice on mental well-being and cognitive performance [2]. Despite these promising reports, novice meditators cannot immediately enjoy meditation benefits as achieving a mindful state is a difficult task and requires intensive training. The challenge lies in the fact that mindfulness meditation is a mental practice which does not yield observable behavioral changes, thus novice meditators cannot receive feedback on their performance from an expert. However, neuroscientific research has shown that brain activity can undergo significant changes during meditation [3, 4] and therefore objective feedback can be provided by assistive systems that monitor users' neurophysiological responses in real-time, helping them sustain the mindful state and avoid mind wandering.

With the recent developments in commercial wearable sensors, there has been a growing interest in neurofeedback-supported personalized meditation training [5, 6]. In most works, the feedback is provided to the user through a non-embodied medium such as audio recordings [5] or immersive VR environments [7]. This is because meditation practice is often considered as a solo activity without any social aspects even though previous reports indicate that meditation in group settings can achieve a higher impact by reducing social anxiety, developing a habit, and helping the group members to reach a preset goal [8, 9]. Similarly, ongoing support and guidance during meditation (e.g. instructions about breathing exercise form a coach) can play a major role in facilitation of practice and regulation of emotions for novice meditators [10]. Thus, it is reasonable to endow meditation assistive systems with a digital mentor, i.e. a social agent, that not only monitors the user's mental state and provides individualized feedback, but also triggers a sense of co-presence and emotional support [11, 12].

The presents study intends to develop a meditative BCI system for administration of mindfulness practice through a social robot. The idea of using a robot guru in this research is inspired by previous HRI studies, proposing that the feeling of co-presence with an embodied therapy agent could enhance motivation and performance while creating a joyful interaction [13, 14]. However, given that brain activity can be influenced by the interaction with a physically present social robot [11] and the induced sense of co-presence [15], we first examined the effect of robot-guided meditation on EEG responses in order to extract an ecologically valid index for neurophysiological changes.

Past electroencephalogram (EEG) studies have introduced multiple features with respect to meditation and mindful state. Many extracted frequency band powers associated with mindfulness practice in novice meditators and reported increased alpha and theta powers mainly in the fronto-central areas of the brain [4, 6, 16]. On the other hand, a few studies have highlighted manifestation of gamma band activity in the parieto-occipital areas as a sign of overall attentive state [17]. Given the abundant number of meditation techniques and individual differences, there are often inconsistent findings among reported spectral activities and hence a consensus is yet to be reached with respect to these features [3].

M. A. is with the Department of Cognitive Science and AI, Tilburg University, Tilburg 5037 AB, Netherlands (corresponding author; e-mail: m.alimardani@uvt.nl).

L. K. was with the Department of Cognitive Science and AI, Tilburg University, Tilburg 5037 AB, Netherlands (e-mail: linda.kemmeren@hotmail.com).

K. O. was with the Graduate School of Arts and Sciences, University of Tokyo, Tokyo 153-0041, Japan.

K. H. is with the Graduate School of Arts and Sciences, University of Tokyo, Tokyo 153-0041, Japan (e-mail: khiraki@idea.c.u-tokyo.ac.jp).

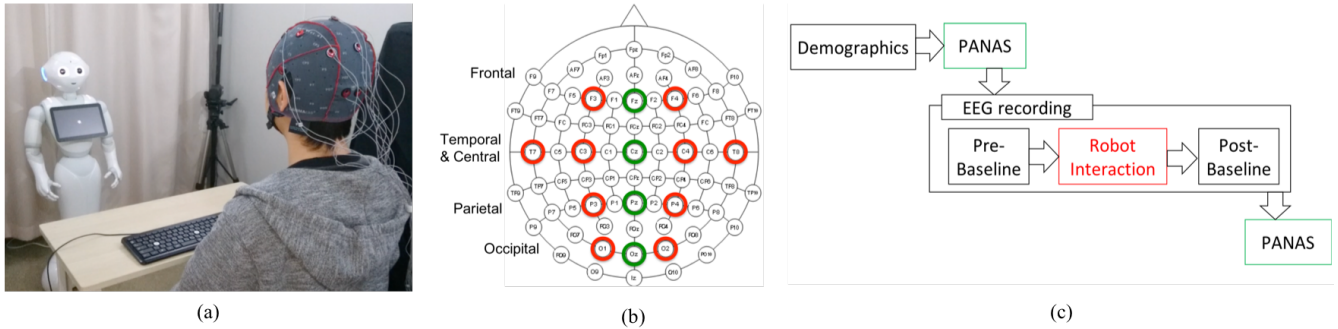

Figure 1. Experimental setup and procedure. (a) Participants listened to a Pepper robot giving them meditation instructions (Meditation group) or a lecture about mindfulness (Control group) while wearing an EEG cap. (b) EEG signals were recorded from 14 channels (all colored electrodes) across five brain regions. Only red electrodes were used in computation of asymmetry power values. (c) The experiment started with a demographic form and PANAS questionnaire, followed by EEG recording in three phases (before, during and after interaction with the robot). Finally, at the end of the experiment, participants filled in PANAS again.

To counter the problem of individual differences and lack of a standard measure for performance evaluation during meditation practice, some studies have employed hemispheric asymmetry analysis (also known as laterality) in order to investigate an individual's brain activity changes in relation to his/her affective state changes before and after meditation. This is in line with the approach-withdrawal theory as introduced by Davidson et al. [18] in which the left frontal region is estimated to be involved in processing of positive emotions, whereas the right frontal region in the negative emotions. Past studies have shown that meditation practice could induce a greater proportion of activity in the left frontal region, particularly in the alpha band, which was associated with positive, approach-oriented emotional states [4, 19, 20]. Similarly, neurofeedback interventions that shift the frontal hemispheric asymmetry toward left activation have been shown to induce improvements in mood and alleviation of anxiety and depression symptoms [21, 22].

In this experiment, we recorded EEG signals from two groups of subjects who either experienced a meditation intervention with a social robot or just listened to the robot talk about mindfulness. In addition to neurophysiological measurements, we measured their self-reported affective state before and after interaction with the robot. We hypothesized that participants in the first HRI condition, who experienced mindfulness meditation with the robot, would demonstrate significant changes in their EEG asymmetry and would experience a positive mood change as measured by the self-assessed affect reports. To the best of our knowledge, this is the first study that combines meditation, EEG and HRI in order to investigate the effect of robot-guided meditation on participants' neurophysiology and affective state.

## II. METHODS

### A. Participants

Twenty-eight participants (mean age = 20.2, SD = 3.4), all novices to meditation, were recruited and divided into two groups. One group was assigned to the Meditation condition and the other to the Control condition (N=14 in each condition). All participants received explanation about the experiment and signed a consent form in accordance with the ethical approval of Ethics Committee of The University of Tokyo.

### B. Materials

For Meditation group, we used a breathing meditation as this was considered an easy practice for beginners. The instruction was given by the Pepper robot (SoftBank robotics) in a natural tone (Fig. 1a). We referred to Williams et al. [23] in preparation of the mindfulness meditative instructions. The Control group listened to a lecture about mindfulness by Pepper, which was also based on this book. The robot used Japanese language in both conditions (as all participants were native Japanese speakers) and conducted non-emotional gestures using hand and head movements.

Brain signals were recorded by a g.USBamp amplifier (g.tec Medical Engineering, Austria) from 14 EEG channels (F3, Fz, F4, T7, C3, Cz, C4, T8, P3, Pz, P4, O1, Oz, and O2) that were placed according to the international 10-20 system (Fig. 1b). Conductive gel was applied to all electrode sites to ensure low-impedance contact between the sensors and scalp. The recorded signals were classified into five regional groups; frontal, central, temporal, parietal, and occipital, covering major lobes of the brain. A reference electrode was mounted on the right ear and a ground electrode on the forehead. All signals were digitally filtered between 0.1-100 Hz and sampled at 1200 Hz.

To measure mood effects, we used the Positive And Negative Affect Scale (PANAS) [24]. The questionnaire consists of 16 questions, answered on a 6-point Likert scale, equally divided in eight items measuring Positive Affect (PA) and eight quantifying Negative Affect (NA). The obtained scores for all eight items in each category were summed in order to compute the total PA and NA scores.

### C. Procedure

After receiving explanation, participants sat in a shielded room where EEG electrodes were placed and the experiment started. The experimental procedure is illustrated in Fig. 1c.

Participants first filled in a demographic form and a pre-interaction PANAS questionnaire. Then, a 1-min baseline EEG was recorded, after which the interaction session with robot started. During this session, which lasted six minutes, the Meditation group practiced mindfulness according to the robot's instructions whereas the Control group only listened to a lecture by the robot about mindfulness. After the session ended, another 1-minute baseline EEG was recorded followed by a post-interaction PANAS questionnaire. Participants closed their eyes during recordings to avoid eye-blink artifacts in the EEG signals. Finally, participants were asked to fill in a questionnaire about their impressions of the experiment and the experiment ended.

*D. Evaluation*

EEG signals were processed in MATLAB. First, quality check reports were created using the FieldTrip toolbox [25] to inspect and reject artifacts. Then, EEGLAB toolbox [26] (v2019.0) was used to band-pass filter the signals and extract mean spectral powers in each experiment phase for five frequency bands; delta (1-3.9 Hz), theta (4-7.9 Hz), alpha (8-12.9 Hz), beta (13-29.9 Hz) and gamma (30-45 Hz). In each frequency band, asymmetry values were computed across all five brain regions by subtracting mean power in the right hemispheric channel from the corresponding left one, i.e. F3-F4 (frontal), C3-C4 (central), T7-T8 (temporal), P3-P4 (parietal) and O1-O2 (occipital). We used these asymmetry values to conduct comparison between the condition groups in two ways; 1) comparison of EEG asymmetry powers during HRI session between the Meditation and Control groups, and 2) comparison of EEG asymmetry change that occurred from resting state pre-baseline to HRI session (ΔAsym) for individuals in the Meditation group vs. Controls. We hypothesized that Meditation group would show a higher frontal asymmetry in either of these comparisons.

Additionally, using PA and NA scores, participants' mood before and after robot interaction were compared across condition groups. Following previous reports on the effect of a single meditation session on mood [27], we hypothesized that participants in the Meditation group would demonstrate a significant increase in their self-assessed PANAS after mindfulness practice with the robot. We also hypothesized that such increase in the subjective measure of PANAS would be in correlation with the observed change in the objective measure of EEG asymmetry from the baseline to HRI session (ΔAsym), particularly in the frontal area.

## III. RESULTS

EEG data from all participants were included in the analysis once artifacts were removed. Nearly all acquired asymmetry values showed a non-normal distribution, therefore a non-parametric Mann-Whitney U test was used for further statistical analysis.

*A. EEG asymmetry during meditative HRI vs. control HRI*

Fig. 2 demonstrates obtained asymmetry values during the robot interaction session in five brain regions (i.e. central, frontal, occipital, parietal, and temporal) and five frequency bands (i.e. delta, theta, alpha, beta and gamma) for the Meditation and Control groups. Asymmetry variations were the highest in the temporal location (particularly in the beta and gamma bands), and the lowest in the frontal, central and parietal locations. In general, most oscillatory activation occurred in the left hemispheric regions, with little differentiation for mindfulness practitioners compared to non-practitioners.

As can be seen in Fig. 2, between-group tests yielded significant difference between Meditation vs. Control groups,

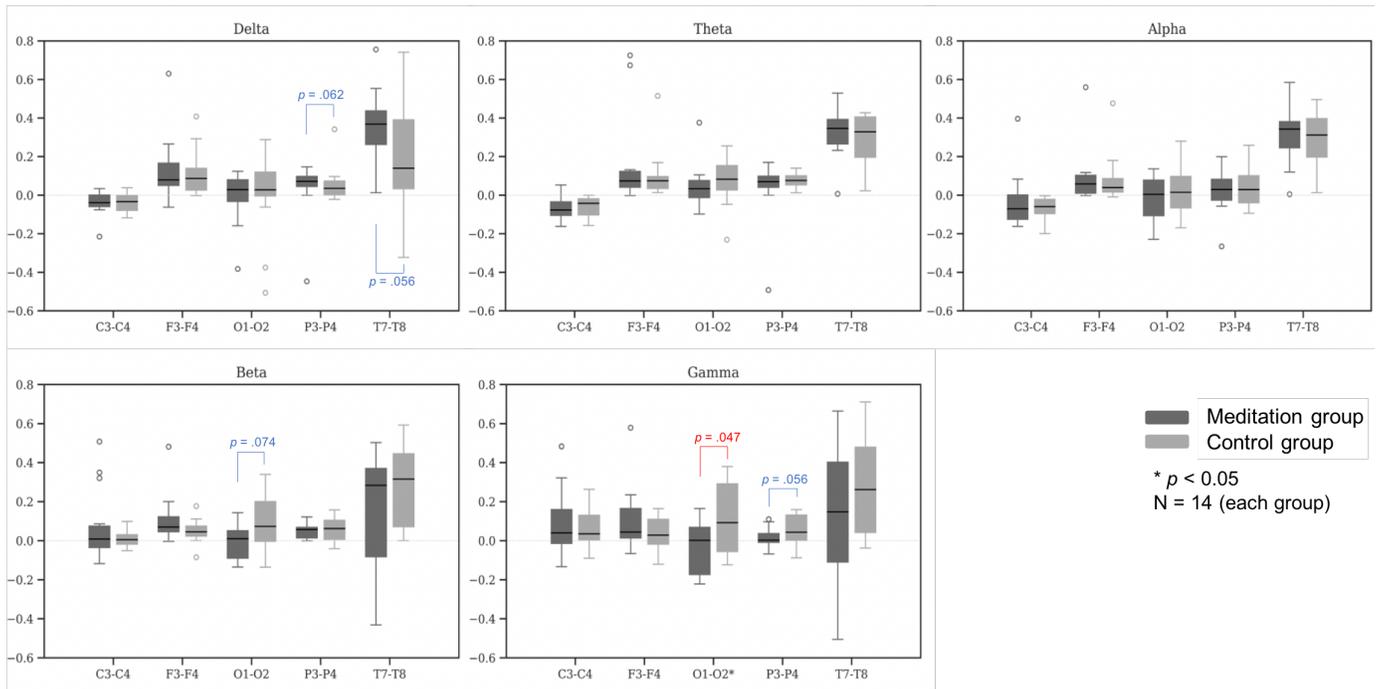

Figure 2. EEG Asymmetry powers during the robot interaction session in five brain regions (i.e. central, frontal, occipital, parietal, and temporal) and five frequency bands (i.e. delta, theta, alpha, beta and gamma) for the Meditation and Control groups.

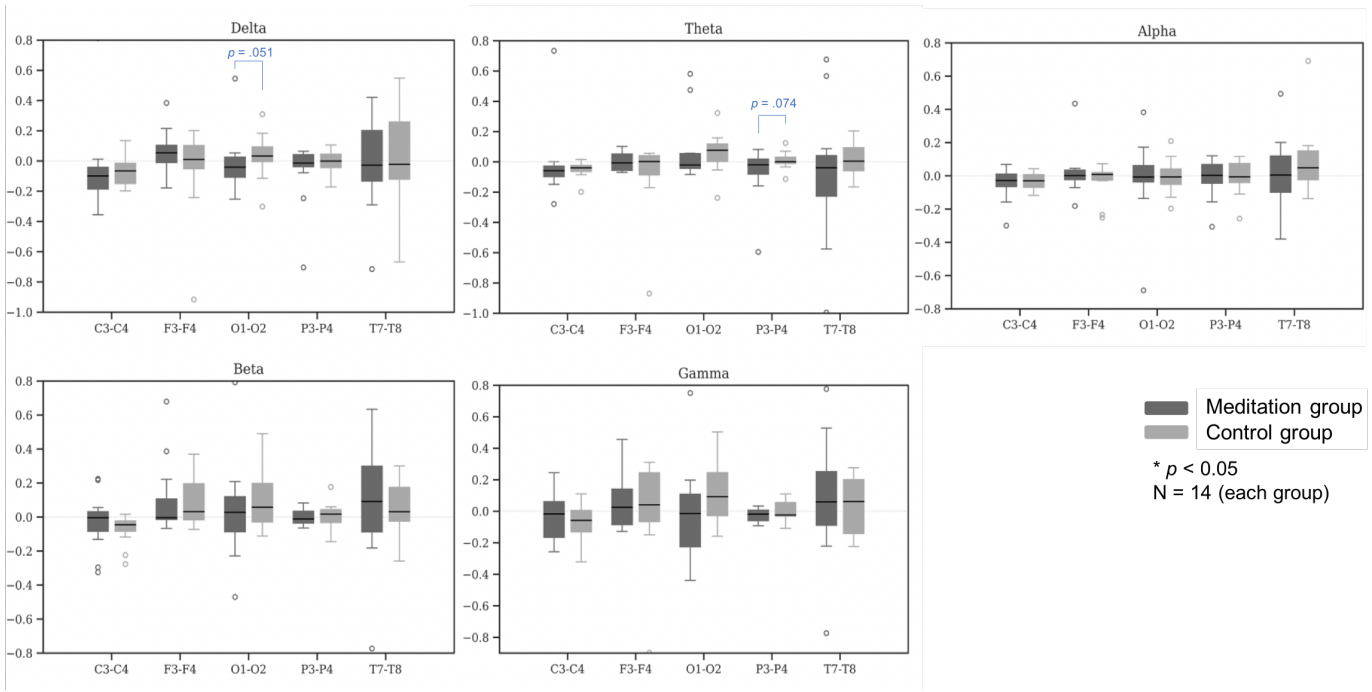

Figure 3. EEG Asymmetry change from the pre-baseline to the robot interaction session in five brain regions (i.e. central, frontal, occipital, parietal, and temporal) and five frequency bands (i.e. delta, theta, alpha, beta and gamma) for the Meditation and Control groups.

only for the occipital gamma asymmetry, which indicated a medium effect size ($p = .047$, $R_u = .378$). Also, trend results were found in occipital beta ($p = .074$), temporal delta ($p = .056$) and in parietal regions in delta ($p = .062$) and gamma bands ($p = .056$). These trend results indicate that differences between participants who practiced mindfulness and those who listened to a lecture on mindfulness during HRI came close to significance level within this sample size, and are therefore marked as trends for future explorations.

### B. Effect of robot-guided meditation on asymmetry change from pre-baseline to HRI session

Fig. 3 visualizes the change in hemispheric asymmetry that participants experienced from resting pre-baseline to robot interaction session (ΔAsym), indicating either a positive change (increase towards left hemispheric asymmetry) or a negative change (decrease towards right asymmetry). Positive or negative ΔAsym values provide no indication of the initial pre-baseline activity and must therefore be interpreted as asymmetry directional change towards left or right activity, rather than exact left or right asymmetry. The figure shows that the asymmetry changes were particularly minor in theta and alpha bands for both groups, with no significant differences. However, large variations in asymmetry change from pre-baseline to HRI session were found in gamma band for almost all locations except parietal. Overall, the graphs indicate that variation in asymmetry change was largest in the temporal site. No clear division between the amount of ΔAsym in Meditation and Control groups was observed in other bands and between-group tests revealed no significant results.

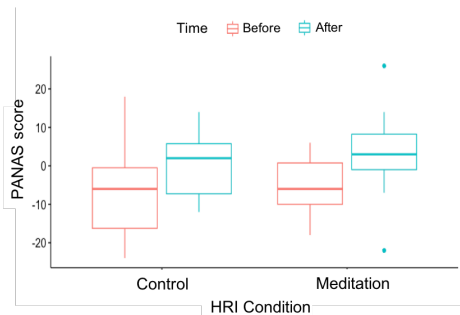

Figure 4. Self-assessed PANAS scores. Participants reported their positive and negative affective state before and after they interacted with the robot.

### C. Self-assessed affective state

The scores for Positive Affect (PA) and Negative Affect (NA) were integrated into one PANAS score by subtracting NA value from PA value. Fig. 4 illustrates obtained PANAS scores aggregated based on the between-group variable 'Group' and within-group variable 'Time'. The Levene's test showed that the variances in the two groups were homogeneous, $F(3, 52) = .461$, $p = .711$. Therefore, a 2x2 factorial ANOVA was conducted on the dataset to evaluate the hypothesis that HRI condition (Group: Meditation vs. Control) and measurement time (Time: before vs. after interaction) would lead to a different score of PANAS. The results did not show a significant main effect for different HRI Group on PANAS scores, $F(1, 52) = .567$, $p = .455$. However, the test showed significant main effect for Time on PANAS scores, $F(1, 52) = 7.978$, $p < .01$, with medium effect size (partial $\eta^2 = .131$). There were no significant interaction effects.

TABLE I. SUMMARY OF SPEARMAN'S CORRELATION COEFFICIENTS

| Frequency band | Brain region | Meditation group | | | Control group | | |
|---|---|---|---|---|---|---|---|
| | | S | p-value | Rho | S | p-value | Rho |
| Delta | Central (C3-C4) | 579.41 | 0.344 | -0.273 | 433.98 | 0.875 | 0.046 |
| | Frontal (F3-F4) | 465.03 | 0.94 | -0.022 | 140.65 | 0.006 | 0.691 |
| | Occipital (O1-O2) | 705.83 | 0.041 | -0.551 | 477.02 | 0.87 | -0.048 |
| | Parietal (P3-P4) | 644.63 | 0.138 | -0.417 | 415.96 | 0.771 | 0.086 |
| | Temporal (T7-T8) | 550.31 | 0.472 | -0.209 | 473.02 | 0.893 | -0.04 |
| Theta | Central (C3-C4) | 381.76 | 0.583 | 0.161 | 304.83 | 0.249 | 0.33 |
| | Frontal (F3-F4) | 391.79 | 0.636 | 0.139 | 261.79 | 0.13 | 0.425 |
| | Occipital (O1-O2) | 536.27 | 0.541 | -0.179 | 355.89 | 0.454 | 0.218 |
| | Parietal (P3-P4) | 470.05 | 0.911 | -0.033 | 287.82 | 0.196 | 0.367 |
| | Temporal (T7-T8) | 382.76 | 0.588 | 0.159 | 493.04 | 0.776 | -0.084 |
| Alpha | Central (C3-C4) | 447.98 | 0.958 | 0.015 | 409.95 | 0.736 | 0.099 |
| | Frontal (F3-F4) | 358.68 | 0.468 | 0.212 | 513.06 | 0.664 | -0.128 |
| | Occipital (O1-O2) | 623.56 | 0.192 | -0.37 | 413.95 | 0.759 | 0.09 |
| | Parietal (P3-P4) | 539.28 | 0.526 | -0.185 | 542.1 | 0.512 | -0.191 |
| | Temporal (T7-T8) | 640.61 | 0.148 | -0.408 | 683.25 | 0.068 | -0.502 |
| Beta | Central (C3-C4) | 504.16 | 0.713 | -0.108 | 501.05 | 0.731 | -0.101 |
| | Frontal (F3-F4) | 429.92 | 0.852 | 0.055 | 394.93 | 0.653 | 0.132 |
| | Occipital (O1-O2) | 618.54 | 0.207 | -0.359 | 350.89 | 0.431 | 0.229 |
| | Parietal (P3-P4) | 601.48 | 0.262 | -0.322 | 326.86 | 0.329 | 0.282 |
| | Temporal (T7-T8) | 472.06 | 0.899 | -0.037 | 535.09 | 0.547 | -0.176 |
| Gamma | Central (C3-C4) | 511.19 | 0.674 | -0.123 | 612.17 | 0.226 | -0.345 |
| | Frontal (F3-F4) | 405.84 | 0.713 | 0.108 | 422.96 | 0.811 | 0.07 |
| | Occipital (O1-O2) | 595.46 | 0.283 | -0.309 | 319.85 | 0.302 | 0.297 |
| | Parietal (P3-P4) | 613.52 | 0.222 | -0.348 | 258.78 | 0.124 | 0.431 |
| | Temporal (T7-T8) | 462.02 | 0.958 | -0.015 | 535.09 | 0.547 | -0.176 |

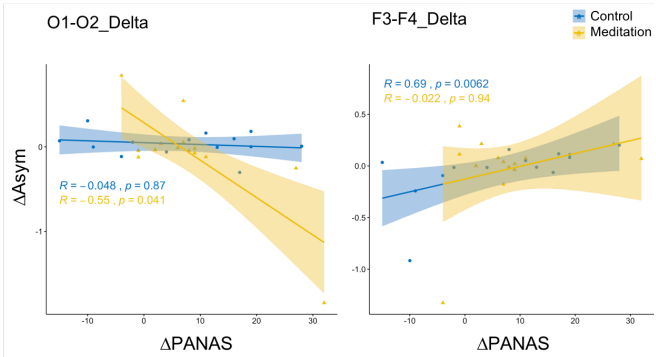

Figure 5. Spearman's correlation between EEG asymmetry change and mood change. Significant correlations were found in occipital delta for Meditation group and in frontal delta for Control group.

Post-hoc analyses for multiple comparisons using Tukey's HSD suggested that none of the results including PANAS scores before Meditation HRI ($M$ = -5.21, $SD$ = 8.20), before Control HRI ($M$ = -6.50, $SD$ = 11.96), after Meditation HRI ($M$ = 3.00, $SD$ = 10.99), and after Control HRI ($M$ = .29, $SD$ = 7.98) showed significant differences.

### D. Correlation between changes in EEG Asymmetry and PANAS scores

The change in participants' mood as quantified by ($\Delta$PANAS = PANAS$_{after}$ − PANAS$_{before}$) was computed for all participants in both groups and paired with the change in EEG asymmetry from pre-baseline to HRI session ($\Delta$Asym). Given the lack of normality in the data, Spearman's rank correlation was conducted to check whether any relationship existed between the two variables $\Delta$Asym and $\Delta$PANAS in each frequency band and each brain region. Table I summarizes the results.

Results of the Spearman correlation indicated that there was a significant negative association between occipital delta asymmetry change and PANAS change in the Meditation group ($r_s$ = -.551, $p$ = .041) and a positive relationship between frontal delta asymmetry change and PANAS change in the Control group ($r_s$ = .691, $p$ = .006) (Fig. 5). Asymmetry changes in other frequency bands and brain areas did not yield any significant correlation with mood changes in participants.

## IV. DISCUSSION

The goal of this research was to explore and evaluate the influence of mindfulness meditation as guided by a social robot on neurophysiological responses such as hemispheric asymmetry, and analyze whether this objective measure correlate with subjective report on perceived mood. The methodology and findings of this study are proposed to contribute to the development of BCI-operated HRI design that support mental health programs and meditation practice through a social robot.

Our results showed a significantly lower EEG asymmetry in the occipital gamma band during robot interaction in the Meditation group compared to the Control group (Fig. 2). The occipital lobe is primarily involved with the processing of visuospatial stimuli, active visualization of images in short-term memory, and selective attention [28-30]. Gamma band activity in this area has been previously associated with the meditative state in experienced meditators and open monitoring of ongoing experience [29, 30]. Cahn and colleagues [29] presented a Vipassana meditation study in which experienced meditators showed higher gamma activation in parieto-occipital areas during closed-eye meditation. The authors attributed their observation to a strong sensory awareness during meditation, a shared characteristic between Vipassana and mindfulness practice. In another study, Berkovich-Ohan et al. [28] identified state increases in parieto-occipital gamma power during mindfulness, largely lateralized to the right, attributed to an increase in attentional skills and open monitoring. Their report is consistent with our results regarding the negative values of occipital gamma asymmetry (lateralization toward right hemispheric activation) that were found in the Meditation group. Given that participants kept their eyes closed during EEG recording and that Meditation group focused on mindful breathing exercises while the Control group only listened to the robot, it is likely that the Control participants were more conscious of the robot presence and experienced a heightened awareness toward the external environment, whereas the Meditation group experienced an internally directed selective attention.

Contrary to the reports of previous studies and our hypothesis in this research, we did not find any significant difference in the frontal asymmetry of the two groups. This can possibly be explained by three reasons; 1) the feeling of co-presence with the robot during interaction session led to a different neurophysiological response than commonly observed in a solo meditation practice, 2) the employed mindfulness practice did not aim at regulation of emotional responses and the EEG asymmetry difference between the two groups was suppressed by the fact that both groups experienced a mild positive mood change after interaction with the robot, and 3) the small sample size and large individual differences in the between-group design.

The sense of co-presence, a construct of social presence in companion robots [31], is defined as the feeling of being with another human or humanlike intelligence and is indispensable for the success and continuous usage of social robots [32]. Past research has shown that the mere presence of another person in the room when participants are involved in a task can induce changes in the brain responses compared to when they conduct the same task alone. For instance, Pozharliev et al. [33] reported significantly higher EEG event-related potentials when participants watched emotionally significant stimuli with another person seated next to them than when they watched the stimuli alone, suggesting that the presence of another person magnified the allocation of attention to the stimuli and emotional processing. Therefore, it could be speculated that the feeling of co-presence with the robot recruited bottom-up attention processes other than the ones associated with meditation, generating more prominent results in the posterior parts of the brain rather than the anterior sites.

The pre- and post-interaction self-reported PANAS scores indicated that almost all participants experienced a relative improvement of mood without any significant main effect between the two groups or significant difference between pre- and post-measurements in each group. The most likely explanation for this outcome is the novelty effect of the HRI setting and the positive attitude that was caused by it after an encounter with the robot. This can be prevented in the future through longitudinal studies that capture the real effect of meditative HRI on participants' affective state once the novelty effect wears off. Another explanation could be that the mindfulness practice employed in this study primarily focused on breathing exercises and was not meant to evoke immediate emotional responses. In previous EEG studies, the mindfulness practice was either conducted for a specific therapy purpose such as anxiety reduction or depression [18, 20] or repeated over a number of weeks and hence participants' mood change before and after that period was estimated [19, 20]. Finally, it could be speculated that the PANAS instrument did not have adequate indicators to capture all affect dimensions. PANAS quantifies a person's negative and positive emotions in an independent way using self-evaluation of words or expressions. As opposed to this, recent studies have employed Self-Assessment Manikin Scale (SAM), which captures mood and emotions in a three-dimensional space (pleasure, activation, and dominance) using universally applicable pictures. Future research can replicate this experiment with a more detailed affect recognition instrument as well as other physiological and behavioral measures that would provide a reliable index for EEG response mapping.

The small sample size (14 participants in each group) was another limitation of this study, which could have led to insignificant results due to individual differences. Although by taking the hemispheric difference, asymmetry measure cancels out some of the EEG signatures associated with individual differences and muscle artifacts, it has been shown that the asymmetry itself can vary across individuals during different cognitive functions such as spatial attention and language processing [34]. In order to reduce the effect of individual differences, alterations in asymmetry powers from pre-baseline to HRI session (ΔAsym) were computed per participant and statistically compared between groups. However, the data did not yield any significant outcome except for a trend in occipital delta activity, with decreased left asymmetry in the Meditation group (Fig. 3). Further correlation analysis indicated a significant negative relationship between ΔAsym in occipital delta and PANAS score changes (Table I, Fig. 5), implying that those who experienced a larger positive mood change after robot-assisted mindfulness, demonstrated a more pronounced decrease is the left hemispheric delta activity. Taking into account that an opposite trend was observed in the delta asymmetry during HRI session in parietal and temporal regions (Fig. 2) as well as a positive correlation in frontal delta of the control group (Fig. 5), it is speculated that these arbitrary slow oscillations were simply a byproduct of the relaxed state that participants experienced during breathing exercises and eye-closure.

Altogether, our findings suggest that neurophysiological measures of brain activity hold promise in detection and online monitoring of users' responses to a robot-assisted therapeutic interventions, although short-term effects are limited and could be confounded by the novelty of the first encounter with the robot. Further research with larger number of participants is required to assess the effect of real-time biofeedback of EEG changes during HRI and long-term impacts of robot-assisted mindfulness practice. Previous research had separately studied brain activity in relation to meditation, emotion or HRI, however, the combination of all is novel to the EEG research. Compared to previous studies, this study was designed upon the assumption that future companion robots can be employed in therapeutic setups and hence investigated how the sense of co-presence with the robot influences EEG responses during a mindfulness intervention. Future research should also investigate the best operations and practices of a robot coach (e.g. monitor performance and provide corrective feedback, offer encouragement when the user needs emotional support, etc.) that would help users achieve the mental and cognitive healthcare benefits that are expected of the intervention [35].

Finally, it is worth noting that in forming of our hypotheses, we identified a gap in the literature with respect to EEG responses during group meditation. More hyper-scanning research is required to explore the effect of co-presence and social motivation on efficient practice of meditation in a group setting, which could lead to comparison of outcomes between robot-guided vs. human-coached meditative interventions. Furthermore, inclusion of more conditions ranging from disembodied voices to humanlike robots and humans can elucidate the effect of co-presence on successful mindfulness therapy and subsequent psychological and cognitive gains for the users.

## V. CONCLUSION

This study presented our first step toward developing a meditation assistive robot and automation of mindfulness therapies. In a between-group experiment, we investigated the EEG correlates of robot-assisted mindfulness practice and their relationships with participants' affective state changes. Our findings provide an interesting result of right occipital gamma asymmetry for mindfulness practitioners, which can be explained by selective attention and open monitoring during mindfulness meditation with the robot.

These results provide new insights for development of neurofeedback systems that monitor a user's brain responses in a social HRI setting and enable the robot to respond to human cognitive performance as well as emotional responses in real-time.

ACKNOWLEDGMENT

This research was supported by ImPACT Program of Council for Science, Technology and Innovation (Cabinet Office, Government of Japan), Grant-in-Aid for JSPS Research Fellow 15F15046, and JST CREST (grant# JPMJCR18A4).

REFERENCES

[1] J. Kabat-Zinn, "Mindfulness-based interventions in context: Past, present, and future," *Clinical Psychology: Science and Practice*, 10(2), 144–156, 2003.
[2] J. Gu, C. Strauss, R. Bond, and K. Cavanagh, "How do mindfulness-based cognitive therapy and mindfulness-based stress reduction improve mental health and wellbeing? A systematic review and meta-analysis of mediation studies," *Clinical psychology review*, 37, 1-12, 2015.
[3] D. J. Lee, E. Kulubya, P. Goldin, A. Goodarzi, and F. Girgis, "Review of the neural oscillations underlying meditation," *Frontiers in neuroscience*, 12, 178, 2018.
[4] T. Lomas, I. Ivtzan, and C. H. Fu, "A systematic review of the neurophysiology of mindfulness on EEG oscillations," *Neuroscience & Biobehavioral Reviews*, 57, 401-410, 2015.
[5] C. Sas, and R. Chopra, "MeditAid: a wearable adaptive neurofeedback-based system for training mindfulness state," *Personal and Ubiquitous Computing*, 19(7), 1169-1182, 2015.
[6] A. A. Fingelkurts, A. A. Fingelkurts, and T. Kallio-Tamminen, "EEG-guided meditation: a personalized approach," *Journal of Physiology-Paris*, 109(4-6), 180-190, 2015.
[7] I. Kosunen, M. Salminen, S. Järvelä, A. Ruonala, N. Ravaja, and G. Jacucci, "RelaWorld: neuroadaptive and immersive virtual reality meditation system," In *Proceedings of the 21st International Conference on Intelligent User Interfaces*, 208-217, March 2016.
[8] Z. Imel, S. Baldwin, K. Bonus, and D. MacCoon, "Beyond the individual: Group effects in mindfulness-based stress reduction," *Psychotherapy Research*, 18(6), 735-742, 2008.
[9] M. Mantzios, and K. Giannou, "Group vs. single mindfulness meditation: exploring avoidance, impulsivity, and weight management in two separate mindfulness meditation settings," *Applied Psychology: Health and Well-Being*, 6(2), 173-191, 2014.
[10] R. Schaub, "Clinical meditation teacher: A new role for health professionals," *Journal of Evidence-Based Complementary & Alternative Medicine*, 16(2), 145-148, 2011.
[11] J. Li, "The benefit of being physically present: A survey of experimental works comparing copresent robots, telepresent robots and virtual agents," *International Journal of Human-Computer Studies*, 77, 23-37, 2015.
[12] A. Shamekhi, and T. Bickmore, "Breathe Deep: A Breath-Sensitive Interactive Meditation Coach," In *Proceedings of the 12th EAI International Conference on Pervasive Computing Technologies for Healthcare*, pp. 108-117, May 2018.
[13] Tapus, A., Tapus, C., & Mataric, M. (2009). The role of physical embodiment of a therapist robot for individuals with cognitive impairments. In *RO-MAN 2009-The 18th IEEE International Symposium on Robot and Human Interactive Communication* (pp. 103-107). IEEE.
[14] Alimardani, M., and Hiraki, K. (2017). Development of a real-time brain-computer interface for interactive robot therapy: an exploration of EEG and EMG features during hypnosis. *Int. J. Comput. Electric. Autom. Control Inform. Eng*, 11, 187-195.
[15] D. Tjon, A. Tinga, M. Alimardani, and M. Louwerse, "Brain Activity Reflects Sense of Presence in 360 degrees Video for Virtual Reality," In *the 28th International Conference on Information Systems Development*, 2019.
[16] C. Kaur, and P. Singh, "EEG derived neuronal dynamics during meditation: progress and challenges" *Advances in preventive medicine*, 2015.
[17] C. Braboszcz, B. R. Cahn, J. Levy, M. Fernandez, and A. Delorme, "Increased gamma brainwave amplitude compared to control in three different meditation traditions," *PLoS One*, 12(1), 2017.
[18] R. J. Davidson, "Asymmetric brain function, affective style, and psychopathology: The role of early experience and plasticity," *Development and Psychopathology*, 6(4), 741-758, 1994.
[19] C. A. Moyer, M. P. Donnelly, J. C. Anderson, K. C. Valek, S. J. Huckaby, D. A. Wiederholt,... and B. L. Rice, "Frontal electroencephalographic asymmetry associated with positive emotion is produced by very brief meditation training," *Psychological science*, 22(10), 1277-1279, 2011.
[20] P. M. Keune, V. Bostanov, M. Hautzinger, and B. Kotchoubey, "Approaching dysphoric mood: state-effects of mindfulness meditation on frontal brain asymmetry," *Biological Psychology*, 93(1), 105-113, 2013.
[21] R. Mennella, E. Patron, and D. Palomba, " Frontal alpha asymmetry neurofeedback for the reduction of negative affect and anxiety," *Behaviour research and therapy*, 92, 32-40, 2017.
[22] J. Tarrant, and H. Cope, "Combining frontal gamma asymmetry neurofeedback with virtual reality: A proof of concept case study," *NeuroRegulation*, 5(2), 57-57, 2018.
[23] M. Williams, J.D. Teasdale, Z. Segal. And J. Kabat-Zinn, "The mindful way through depression: Freeing yourself from chronic unhappiness," Guilford Press, 2007.
[24] D. Watson, L. A. Clark, A. Tellegen, "Development and validation of brief measures of positive and negative affect: the PANAS scales," *Journal of personality and social psychology*, 54(6), 1063, 1988.
[25] R. Oostenveld, P. Fries, E. Maris, and J. M. Schoffelen, "FieldTrip: open source software for advanced analysis of MEG, EEG, and invasive electrophysiological data," *Computational intelligence and neuroscience*, 2011.
[26] A. Delorme, and S. Makeig, "EEGLAB: an open source toolbox for analysis of single-trial EEG dynamics including independent component analysis," *Journal of neuroscience methods*, 134(1), 9-21, 2004.
[27] S. Johnson, R. M. Gur, Z. David, and E. Currier, "One-session mindfulness meditation: a randomized controlled study of effects on cognition and mood," *Mindfulness*, 6(1), 88-98, 2015.
[28] A. Berkovich-Ohana, J. Glicksohn, and A. Goldstein, "Mindfulness-induced changes in gamma band activity–implications for the default mode network, self-reference and attention," *Clinical neurophysiology*, 123(4), 700-710, 2012.
[29] B. R. Cahn, A. Delorme, and J. Polich, "Occipital gamma activation during Vipassana meditation," *Cognitive processing*, 11(1), 39-56, 2010
[30] F. Travis, and J. Shear, "Focused attention, open monitoring and automatic self-transcending: categories to organize meditations from Vedic, Buddhist and Chinese traditions," *Consciousness and cognition*, 19(4), 1110-1118, 2010.
[31] I. Leite, C. Martinho, A. Pereira, and A. Paiva, "As time goes by: Long-term evaluation of social presence in robotic companions," In *RO-MAN 2009-The 18th IEEE International Symposium on Robot and Human Interactive Communication* (pp. 669-674). IEEE.
[32] K. M. Lee, W. Peng, S. A. Jin, and C. Yan, "Can robots manifest personality?: An empirical test of personality recognition, social responses, and social presence in human–robot interaction," *Journal of communication*, 56(4), 754-772, 2006.
[33] R. Pozharliev, W. J. Verbeke, J. W. Van Strien, and R. P. Bagozzi, "Merely being with you increases my attention to luxury products: Using EEG to understand consumers' emotional experience with luxury branded products," *Journal of Marketing Research*, 52(4), 546-558, 2015.
[34] E. Ambrosini, and A. Vallesi, "Asymmetry in prefrontal resting-state EEG spectral power underlies individual differences in phasic and sustained cognitive control," *Neuroimage*, 124, 843-857, 2016.
[35] S. M. Rabbitt, A. E. Kazdin, and B. Scassellati, "Integrating socially assistive robotics into mental healthcare interventions: Applications and recommendations for expanded use," *Clinical psychology review*, 35, 35-46, 2015.